# Satellite-to-ground quantum key distribution

Sheng-Kai Liao[1,2], Wen-Qi Cai[1,2], Wei-Yue Liu[1,2], Liang Zhang[2,3], Yang Li[1,2], Ji-Gang Ren[1,2], Juan Yin[1,2], Qi Shen[1,2], Yuan Cao[1,2], Zheng-Ping Li[1,2], Feng-Zhi Li[1,2], Xia-Wei Chen[1,2], Li-Hua Sun[1,2], Jian-Jun Jia[3], Jin-Cai Wu[3], Xiao-Jun Jiang[4], Jian-Feng Wang[4], Yong-Mei Huang[5], Qiang Wang[5], Yi-Lin Zhou[6], Lei Deng[6], Tao Xi[7], Lu Ma[8], Tai Hu[9], Qiang Zhang[1,2], Yu-Ao Chen[1,2], Nai-Le Liu[1,2], Xiang-Bin Wang[2], Zhen-Cai Zhu[6], Chao-Yang Lu[1,2], Rong Shu[2,3], Cheng-Zhi Peng[1,2], Jian-Yu Wang[2,3], Jian-Wei Pan[1,2]

[1] Shanghai Branch, Department of Modern Physics and National Laboratory for Physical Sciences at the Microscale, University of Science and Technology of China, Shanghai 201315, China .
[2] CAS Center for Excellence and Synergetic Innovation Center in Quantum Information and Quantum Physics, University of Science and Technology of China, Shanghai 201315, China
[3] Key Laboratory of Space Active Opto-Electronic Technology, Shanghai Institute of Technical Physics, Chinese Academy of Sciences, Shanghai 200083, China.
[4] National Astronomical Observatories, Chinese Academy of Sciences, Beijing 100012, China
[5] Key Laboratory of optical engineering, The Institute of Optics and Electronics, Chinese Academy of Sciences, Chengdu 610209, China
[6] Shanghai Engineering Center for Microsatellites, Shanghai 201203, China
[7] State Key Laboratory of Astronautic Dynamics, Xi'an Satellite Control Center, Xi'an 710061, China
[8] Xinjiang Astronomical Observatory, Urumqi 830011, China
[9] National Space Science Center, Chinese Academy of Sciences, Beijing 100190, China

**Abstract:**

Quantum key distribution (QKD) uses individual light quanta in quantum superposition states to guarantee unconditional communication security between distant parties. In practice, the achievable distance for QKD has been limited to a few hundred kilometers, due to the channel loss of fibers or terrestrial free space that exponentially reduced the photon rate. Satellite-based QKD promises to establish a global-scale quantum network by exploiting the negligible photon loss and decoherence in the empty out space. Here, we develop and launch a low-Earth-orbit satellite to implement decoy-state QKD with over kHz key rate from the satellite to ground over a distance up to 1200 km, which is up to 20 orders of magnitudes more efficient than that expected using an optical fiber (with 0.2 dB/km loss) of the same length. The establishment of a reliable and efficient space-to-ground link for faithful quantum state transmission constitutes a key milestone for global-scale quantum networks.

**Introduction**

Private and secure communications are fundamental human needs. Traditional public key cryptography usually relies on the perceived computational intractability of certain mathematical functions. In contrast, quantum key distribution (QKD)[1] proposed in the mid-1980s—the best known example of quantum cryptographic tasks—is a radical new way to offer an information-theoretically secure solution to the key exchange problem, ensured by the laws of quantum physics. QKD allows two distant users, who do not share a long secret key initially, to produce a common, random string of secret bits, called a secret key. Using the one-time pad encryption, this key is proven to be secure by Shannon[2] to encrypt (and decrypt) a message, which can then be transmitted over a standard communication channel. In the QKD, the information is encoded in the superposition states of physical carriers at single-quantum level, where photons, the fastest flying qubits with their intrinsic robustness to decoherence and ease of control, are usually used. Any eavesdropper on the quantum channel attempting to gain information of the key will inevitably introduce disturbance to the system, and can be detected by the communicating users.

Since the first table-top QKD experiment[3] in 1989 with a quantum channel distance of 32 cm, a strong research effort has been devoted to achieve secure QKD at long distance, eventually aiming at global scale for practical use. The most straightforward method is directly sending single photons through optical fibers or terrestrial free-space. In both cases, however, the channel loss cause a decrease of the transmitted photons that scales exponentially with the length. Unlike classical telecommunications, the quantum signal in QKD cannot be noiselessly amplified due to the quantum non-cloning theorem[4]. This limits the maximal distance for secure QKD to a few hundred kilometers[5]. Beyond this length scale, quantum communications become extremely challenging[6].

To overcome this problem, one solution is to employ quantum repeaters[7] that combine entanglement swapping[8], entanglement purification[9], and quantum memories[10]. In spite of remarkable progress in the demonstrations of the three building blocks[11–13] and even prototype quantum repeater nodes[14-18], these laboratory technologies are still far from being practically applicable in realistic long-distance quantum communications.

A more direct and promising solution for global-scale QKD is through satellites in space. Compared with terrestrial channels, the satellite-to-ground connection has significantly reduced losses[19]. This is mainly because that the effective thickness of the atmosphere is ~10 km, and most of the photon's propagation path is in empty space with negligible absorption and turbulence. A ground test[20] in 2004 has demonstrated the distribution of entangled photon pairs over a noisy ground atmosphere of 13 km—beyond the effective thickness of the aerosphere—and showed the survival of entanglement and violation of Bell's inequality. Further verifications of the feasibilities of the satellite-based QKD, under the simulated conditions of huge attenuation and various turbulence, have been performed at even longer distance[21–23], on rapidly moving platforms[24,25], and exploiting satellite corner cube retroreflectors[26,27].

We have developed a sophisticated satellite, named after *Micius*, dedicated for quantum science experiments (for the project timeline and its design details, see Methods), which was successfully launched on 16th August 2016, from Jiuquan, China, orbiting at an altitude of ~500 km (Fig. 1a). Using one of the satellite payloads—a decoy-state QKD transmitter at 850 nm wavelength—and cooperating with Xinglong ground observatory station (near Beijing, 40°23'45.12''N, 117°34'38.85''E, altitude 890m), we establish the decoy-state QKD with polarization encoding from the satellite to the ground with ~kHz rate over a distance up to 1200 km.

**Experimental challenges and solutions**

A robust and efficient satellite-to-ground QKD places a more stringent requirement on the link efficiency than conventional satellite-based classical communication systems. To obtain a high signal-to-noise ratio, one cannot increase the signal power, but only reduce the channel attenuation and background noise. In our experiment, several effects contribute to channel loss, including beam diffraction, pointing error, atmospheric turbulence and absorption.

In our QKD experiment, we adopt the downlink protocol—from the satellite to ground (see Fig. 1a). In the downlink, beam wandering caused by the atmospheric turbulence occurs in the very end of the transmission path (near the earth surface), where the beam size due to diffraction is typically much larger than the beam wandering. Therefore, the downlink has reduced beam spreading compared to the uplink and thus has higher link efficiency.

The beam diffraction mainly depends on telescope size. To narrow the beam divergence, we use a 300-mm aperture Cassegrain telescope in the satellite (Fig. 1b) optimized to eliminate chromatic and spherical aberrations, which sends the light beam with a near-diffraction-limited far-field divergence of ~10 μrad. After a travel distance of 1200 km, we expect the beam diameter expands to about 10 m. At the ground station, a Ritchey-Chretien telescope with an aperture of 1 m and a focal length of 10 m (Fig. 1c) is used to receive the QKD photons (see Methods). The diffraction loss is estimated to be 22 dB at 1200 km.

The narrow divergence beam from the fast-moving satellite (with a speed of ~7.6 km/s) demands a high-bandwidth and high-precision acquiring, pointing, and tracking (APT) system to establish a stable link. We design cascaded multi-stage APT systems in the transmitter (Fig. 1b) and the receiver (Fig. 1c). Initial coarse orientation of the telescope is based on forecasted satellite orbit position with an uncertainty below 200 m. The satellite attitude control system itself ensures the transmitter pointing to the ground station with ~0.5° precision. The satellite and the ground station send beacon lasers to each other with a divergence of 1.25 mrad and 0.9 mrad, respectively (Fig. 2a). The coarse pointing stage in the satellite transmitter consists of a two-axis gimbal mirror (with a range of 10° in both azimuth and elevation) and a CMOS camera with a field-of-view of 2.3°×2.3° and frame rates of 40 Hz. The fine pointing stage uses a fast steering mirror driven by piezo ceramics (with a tracking range of 1.6 mrad) and a camera with a field-of-view of 0.64 mrad×0.64 mrad and frame rates of 2 kHz. Similar coarse and fine APT systems are also equipped in the ground station (see Extended Data Table 2 for details). Using a feedback closed-loop, the transmitter achieves a tracking accuracy of ~1.2 μrad (see Fig. 2b), much smaller than the beam divergence. We estimate that at 1200 km the loss due to atmospheric absorption and turbulence is in the range of 3 dB to 8 dB, and the loss due to pointing error is less than 3 dB.

Furthermore, we use temporal and spectral filtering to suppress the background noise. The beacon laser, with a 0.9-ns pulse width and a ~10-kHz repetition rate, serves for both the APT and synchronization. In a good co-alignment with the QKD photons, the beacon laser can be separated by a dichroic mirror and detected by a single-photon detector in the ground station for timing information. Thus, we avoid the space-ground clock drift, and obtain a synchronization jitter of 0.529 ns, which is used to tag the received signal photons within a 2-ns time window and filter out the background noise. In addition, spectrally, we use a bandwidth filter in the receiver to reduce the background scattering. In the current experiment, we limit ourselves to night-time operation only to avoid sun light.

Finally, we note that the relative motion of the satellite and the ground station induces a time-dependent rotation of the photon polarization seen by the receiver. During one orbit, theoretically we can predict that the polarization contrast ratio would drop from 150:1 to 0 (Fig. 2c). To solve this problem, we calculate rotation angle offset by taking into account of their relative motion and all the birefringent elements in the optical path. Using a motorized HWP for dynamical polarization compensation during the satellite passage, the average polarization contrast ratio increases to 280:1, as shown in Fig. 2c.

**Experimental procedure and results**

In our experiment, we use the decoy-state[28,29] Bennett-Brassard 1984 (BB84)[1] protocol for the QKD, which can detect photon-number-splitting eavesdropping and thus allow secure QKD using weak coherent pulses with significantly increased distance and rate. The key method is to use multiple intensity levels at the transmitter's source, one signal state (mean photon number $\mu_s$) and several decoy states ($\mu_1$, $\mu_2$, …) which are randomly interspersed. Here we adopt a 3-intensity protocol using three levels of $\mu$: a high $\mu_s$, a moderate $\mu_1$ and a zero $\mu_2$ (vacuum), sent with probabilities of 50%, 25%, and 25%, respectively, which are optimized by performing simulations to maximize the secret bit rate for the satellite-to-ground channel.

For downlink QKD, a space-qualified transmitter is integrated in the satellite (see Fig. 1b). Eight fiber-based laser diodes—four used as signal and four as decoy state—emit laser pulses (848.6 nm, 100 MHz, 0.2 ns). The output power of the eight laser didoes are monitored in real time by internal integrated photodetectors and remotely controlled by closed-loop systems, which precisely set the required intensity of the signal and decoy states and stabilize with less than 5% variation. In-orbit measurements show that with independent temperature tuning of the eight lasers, their wavelengths are matched within 0.006 nm, much smaller than their intrinsic bandwidth (~0.1 nm). The lasers are synchronized to be within <10 ps, much smaller than their pulse duration of ~200 ps. The output beams are coaligned to ensure that both concentricity and coaxiality are better than 95%.

The light beams are then sent to a BB84-encoding module consisting of a half-wave plate (HWP), two polarizing beam splitters (PBSs), and a beam splitter (BS), which randomly prepares the emitted photons in one of the four polarization states: horizontal, vertical, linear +45° and -45°. A physical thermal noise device generates a 4-bit random number for each run that drives the eight lasers and determines the output polarization and intensity levels. Independent electric control of the eight lasers and adjustment of the attenuation allow us to accurately obtain the average photon number in the output of the telescope: $\mu_s$=0.8, $\mu_1$=0.1, $\mu_2$=0. In the ground station, a compact decoding setup consisting of a BS, two PBSs, and four single-photon detectors (efficiency 50%, dark counts <25 Hz, timing jitter 350 ps) are used for polarization state analysis (see Fig. 1c and Methods). The overall optical efficiency including the receiving telescope and the fiber coupling on the ground station is ~16%. The satellite uses radio frequency channel for classical communication with the ground station (with an uplink and downlink bandwidth of 1 Mbps and 4 Mbps, respectively), and exploits its experimental control box payload to perform the sifting, error correction and privacy amplification.

The satellite passes Xinglong ground station along a sun-synchronous orbit once every night starting at around 12:50 PM local time, for a duration of about 5 minutes. About 10 minutes before the satellite enters the shadow zone, its attitude is adjusted to point at the ground station. When the satellite exceeds an elevation angle of 5° from the ground station's horizon plane, a pointing accuracy of better than 0.5° is achieved. Then the APT systems start bidirectional tracking and pointing to guarantee that the transmitter and receiver are robustly locked through the whole orbit. From about 15° elevation angle, the QKD transmitter sends randomly modulated signal and decoy photons, together with the beacon laser for timing synchronization, which are received and detected by the ground station. A single-orbit experiment ends when the satellite reaches 10° elevation angle in the other end (see Methods).

Since September 2016, we have been able to successfully perform QKD routinely under good atmospheric condition. Figure 3a shows the data for the orbit on 19th December 2016 with the minimal (maximal) separation of 645 km (1200 km). Within a duration of 273 s for the QKD data collection, the ground station collected 3,551,136 detection events, and thereof 1,671,072 bits of sifted keys (see Fig. 3b). The sifted key rate decrease from ~12 kbit/s at 645 km to ~1 kbit/s at 1200 km, because of the increase of both the physically separated distance and the effective thickness of the atmosphere near the earth at smaller elevation angles. The time trace of the sifted key rate in Fig. 3b demonstrates that we are able to obtain the keys through the whole duration reliably. We note, however, more pronounced key rate fluctuation is observed in the central points when the satellite passes through the ground station around the top, where its effective angular velocity reaches maximum, ~1°/s, thus placing stringent demand on the APT system. Figure 3c shows the observed quantum bit error rate (QBER) with an average of 1.1%, consistent with the expected error rate due to background noise and polarization visibility. The QBERs become slightly higher in the second half of the orbit when the ground telescope points to Beijing that brings more city stray light.

We then perform error correction and privacy amplification to obtain final keys. After randomly shuffling the sifted key, a hamming algorithm is used for error correction. Further, we perform privacy amplification to reduce Eve's possible knowledge by applying random matrix over the corrected keys. Moreover, we take into account of the intensity fluctuation for the signal state and decoy state (<5%), and we calculate secure final key of 300,939 bits when the statistical failure probability is set to be $10^{-9}$, corresponding to a key rate of ~1.1 kbit/s. As in previous experiments[24,25], here the key analysis doesn't consider information leakage due to possible side channels such as the imperfect spatial, temporal and spectral overlap of the quantum light sources.

The QKD experiments performed at 23 different days is summarized in Extended Data Table 3 and Extended Data Figure 9, where the physical distance between the satellite and the ground station varies for different days. The shortest satellite-to-station distance depends on the highest altitude angle of the day, which varies from 507.0 km at 85.7° to 1034.7 km at 25.0°. The obtained sifted key has a peak rate of 40.2 kbits/s at 530 km and decreases at larger distances, for instance, to 1.2 kbits/s at 1034.7 km. From Extended Data Figure 9, we also observe the key rate fluctuation due to different weather conditions. The QBERs are measured to be in the 1%-3% range.

We compare the performance of our satellite-based QKD with what expected from the conventional method of direct transmission through telecommunication fibers. Figure 4 shows the extracted link efficiency at the distance from 645 km to 1200 km from the observed count rate, together with theoretically calculated link efficiency using fibers with 0.2 dB/km loss. Despite the short coverage time (273 s per day) using the *Micius* satellite and the need for reasonably good weather condition, an increasing efficiency enhancement is pronounced at long distances. At 1200 km, the satellite-based QKD within the 273 s coverage time demonstrates a channel efficiency that is ~20 orders of magnitudes higher than using the optical fiber. As a comparison with our data in Fig. 3b, through a 1200 km fiber, even with a perfect 10-GHz single-photon source and ideal single-photon detectors with no dark count, one would obtain only 1-bit sifted key over six million years.

**Discussion and outlook**

We have reported the first satellite-to-ground quantum communication experiment over 1200 km distance scale. Our satellite can be further exploited as a trustful relay to conveniently connect any two points on the earth for high-security key exchange. For example, we can first implement QKD in Xinglong, after which the key is stored in the satellite for 2 hours until it reaches Nanshan station near Urumqi, by a distance of ~2500 km from Beijing. By performing another QKD between the satellite and the Nanshan station, and using one-time-pad encoding, secure key between Xinglong and Nanshan can then be established. Future experimental plan also includes intercontinental secure key exchanges between China and Austria, Italy, and Germany.

Thus far, the low-Earth-orbit satellite has shortcomings of limited coverage area and amount of time spent in each ground station. To increase the coverage, we plan to launch satellites at higher orbit and construct a satellite constellation, which require the development of new techniques to increase the link efficiency, including larger-size telescopes, better APT systems, and wave-front correction through adaptive optics. Higher-orbit satellites, however, will spend less time in the earth's shadow. Day-time QKD can be implemented using telecommunication wavelength photons and improved spatial and spectral filtering[30].

The satellite-based QKD can be linked to metropolitan quantum networks where fibers are sufficient and convenient to connect numerous users within a city at ~100 km scale[31].

We can thus envision a space-ground integrated quantum network, enabling quantum cryptography—most likely the first commercial application of quantum information—useful at a global scale.

**Figure captions:**

**Figure 1 | Illustration of the experimental set-up. a,** Overview of the satellite-to-ground QKD. The *Micuis* satellite, weighted 635 kg, flies along a sun-synchronous orbit at an altitude of ~500 km. It is equipped with three space-qualified payloads to accomplish a series of space-scale quantum experiments including QKD, Bell test, and teleportation. **b,** Schematic of the decoy-state QKD transmitter which is one of the satellite payloads. Attenuated laser pulses (~850 nm) from eight separate laser diodes (LD1, ⋯, LD8) pass through a BB84 encoding module (that consists of two PBSs, a HWP and a BS), co-aligned with a green laser beam (LA1) for system tracking and time synchronization, and is sent out through a 300 mm aperture Cassegrain telescope. After the BB84 module, a ~5 μW laser is used as a polarization reference. A two-axis gimbal mirror (GM1) in the output of the telescope and a large field-of-view camera (CAM1) are combined for coarse tracking loop control. Two fast steering mirrors (FSM) and a fast camera (CAM2) are used for fine tracking. **c,** Schematic of the decoy-state QKD decoder in the Xinglong ground station that equipped with a 1000-mm-aperture telescope. The received 532 nm laser is separated by a dichromic mirror (DM) and split into two paths: one is imaged by a camera (CAM3) for tracking, and the other one is detected for time synchronization. The 850 nm decoy-state photons are analyzed by a BB84 decoder that consists of a BS and two PBS, and detected by four single-photon detectors (SPDs). The ground station sends a red laser (LA2) beam to the satellite for system tracking. See Table I for more technical parameters.

**Figure 2 | Establishment of a reliable space-to-ground link for quantum state transfer. a**, Overlaid and time-lapse photographs of tracking laser beams as the satellite flies over the Xinglong station. The red and green lasers are sent from the ground and

the satellite, respectively, with a divergence of 1.2 mrad. **b,** Long-time tracking error of both X and Y axis extracted from the real-time images read out from the fast camera. **c,** Polarization contrast ratio with and without dynamical compensation during one orbit.

**Figure 3 | Performance of satellite-to-ground QKD performance during one orbit. a,** The trajectory of the *Micuis* satellite measured from Xinglong ground station. **b,** The sifted key rate as a function of time and physical distance from the satellite to the station. **c,** Observed quantum bit error rate. See text for detailed discussions on the results and see Extended Data Table 2 and Extended Data Figure 8 for additional data on different days.

**Figure 4 | A comparison of the QKD link efficiencies between direct transmission through 0.2 dB/km-loss telecommunication optical fibers (red points) and the satellite-to-ground approach (blue points).** The link efficiency was calculated by dividing the photon intensity arrived in front of the detectors in the receiving ground station by that at the satellite transmitter's output. At a distance of 1200 km, the latter (within the satellite coverage time) is more efficient than the former by 20 orders of magnitude.

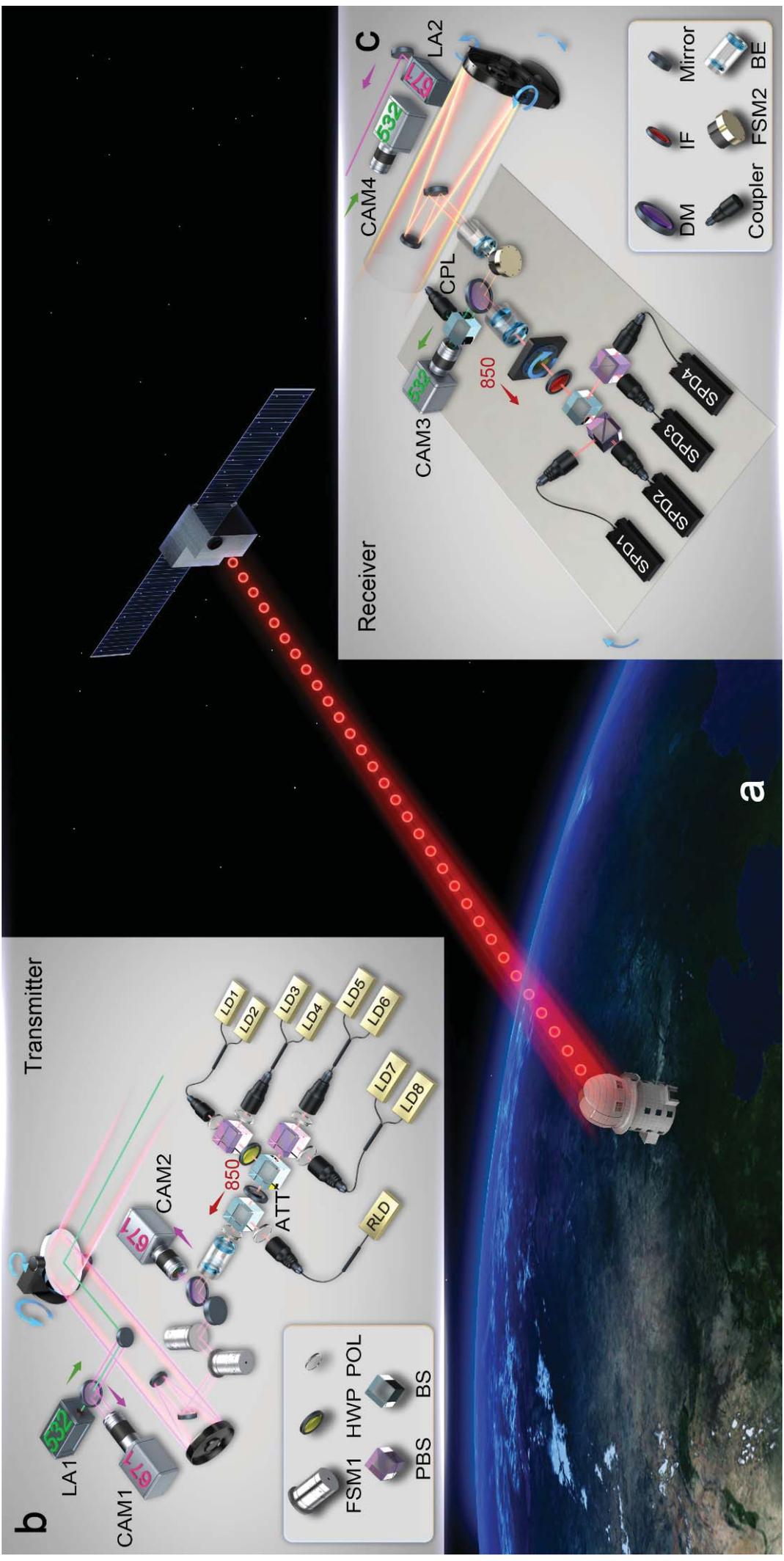

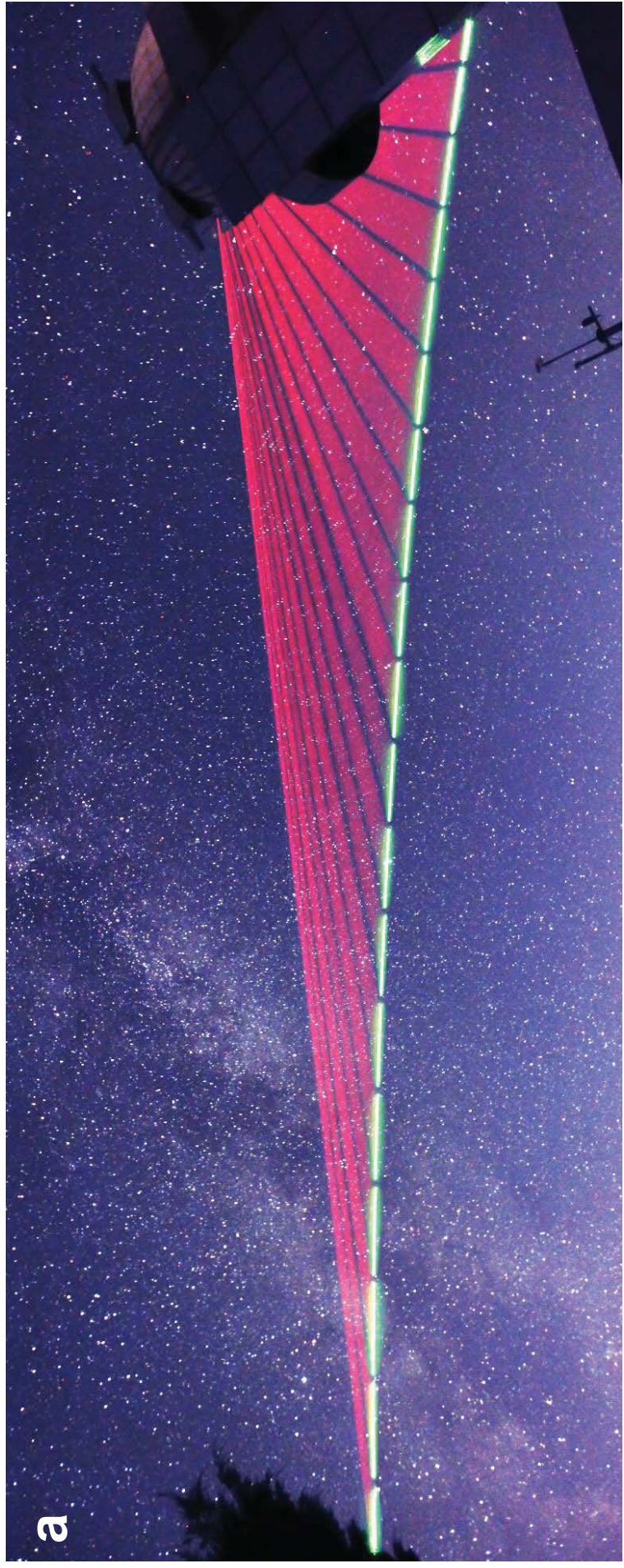

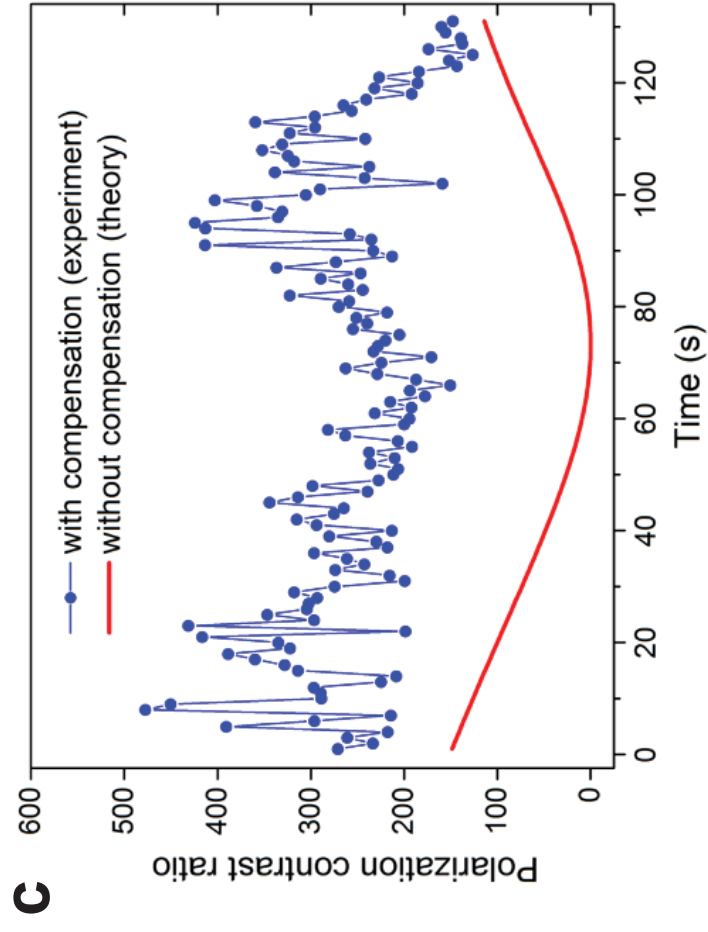

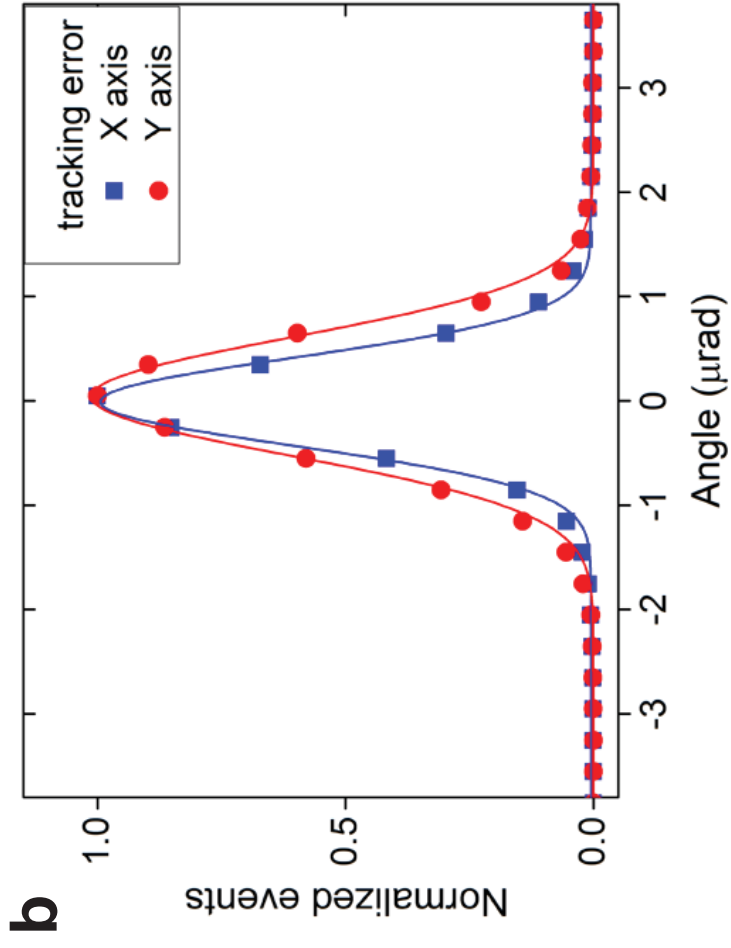

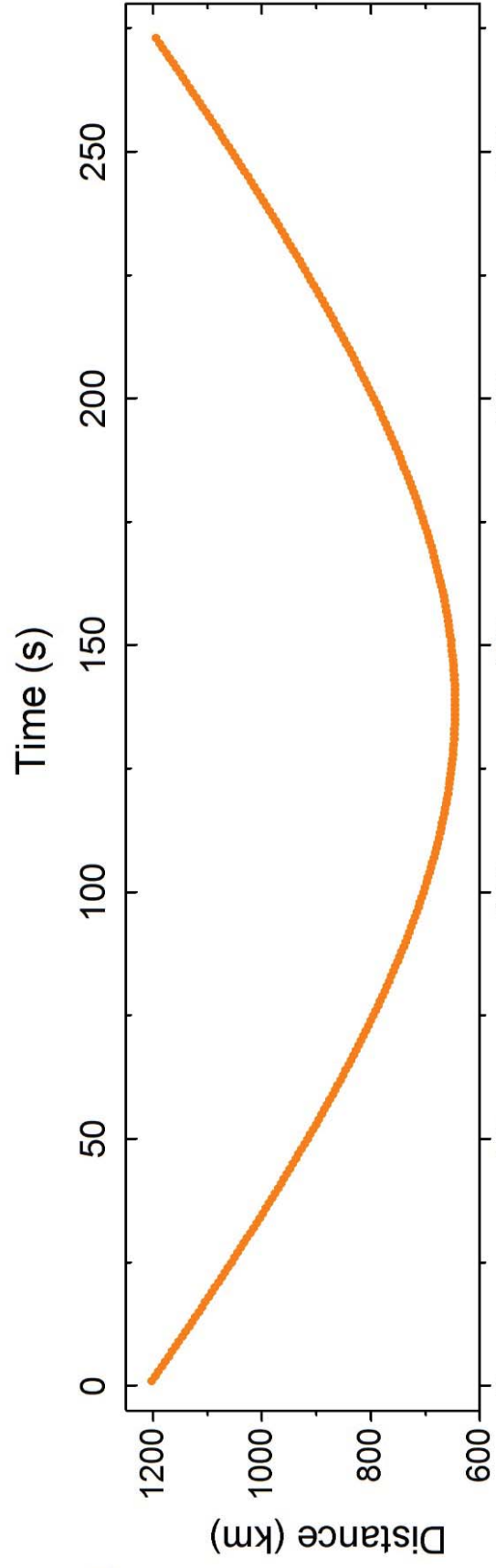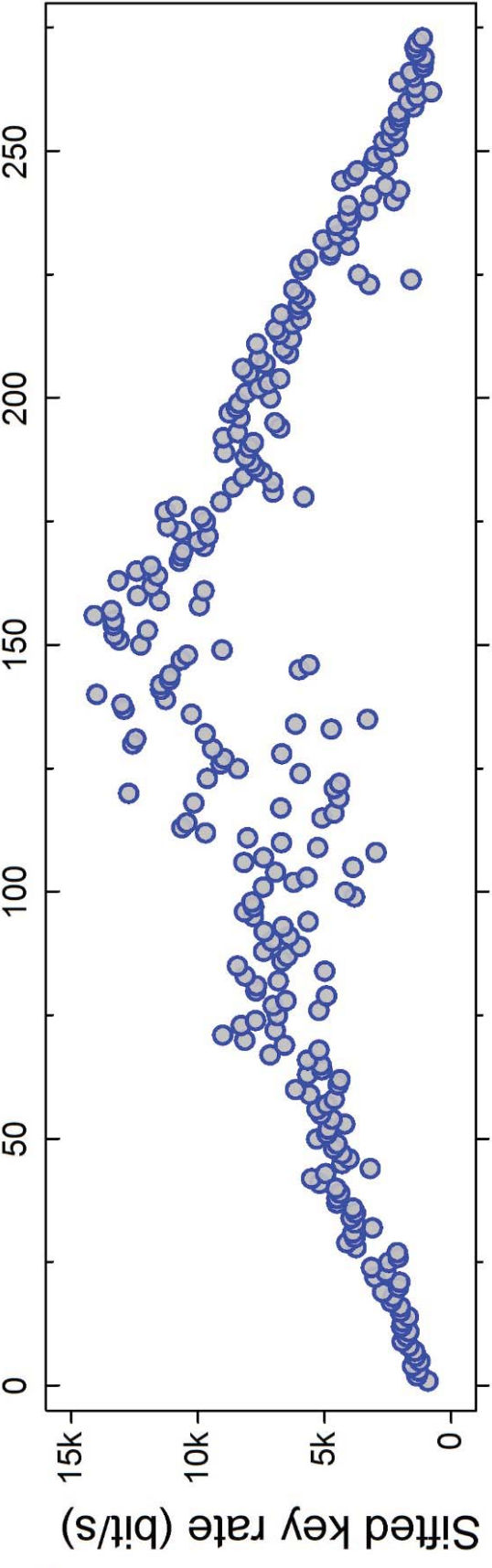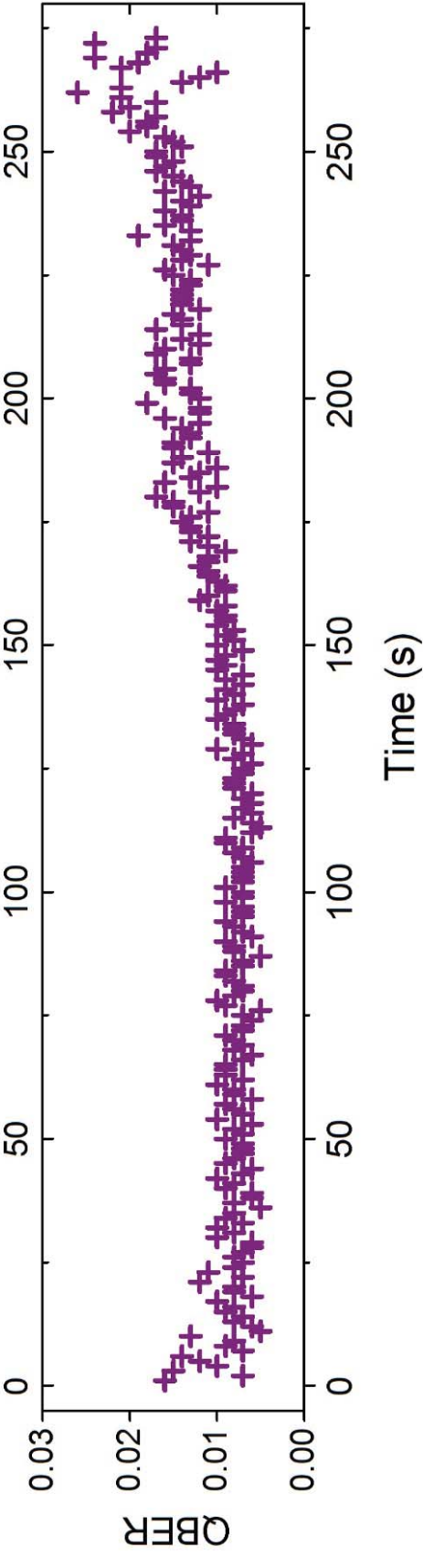

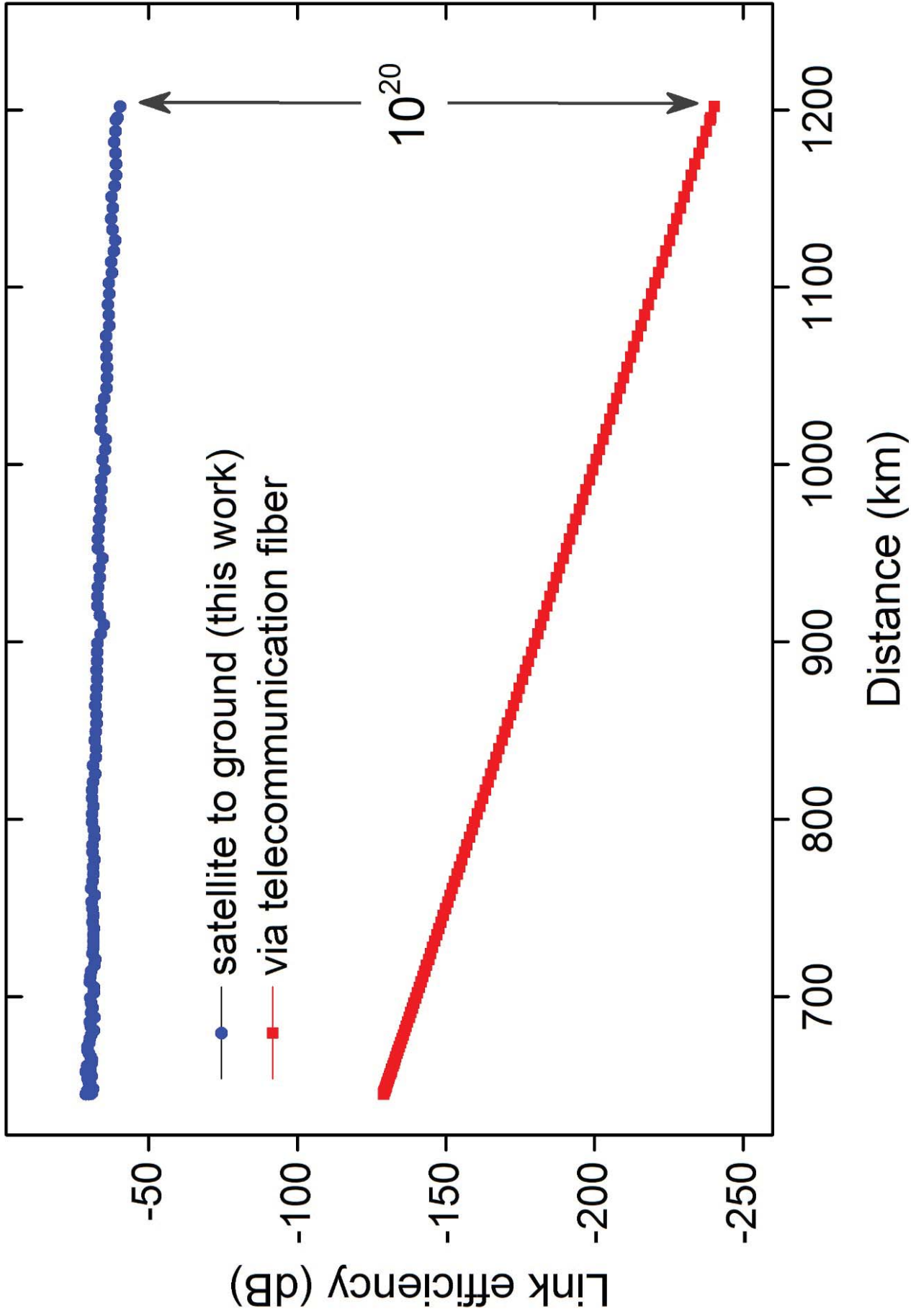